# FOURTH-ORDER GRAVITY GRADIENT TORQUE OF SPACECRAFT ORBITING ASTEROIDS[*]

Yue Wang,[†] Hong Guan,[‡] and Shijie Xu[§]

The dynamical behavior of spacecraft around asteroids is a key element in design of such missions. An asteroid's irregular shape, non-spherical mass distribution and its rotational sate make the dynamics of spacecraft quite complex. This paper focuses on the gravity gradient torque of spacecraft around non-spherical asteroids. The gravity field of the asteroid is approximated as a 2nd degree and order-gravity field with harmonic coefficients $C_{20}$ and $C_{22}$. By introducing the spacecraft's higher-order inertia integrals, a full fourth-order gravity gradient torque model of the spacecraft is established through the gravitational potential derivatives. Our full fourth-order model is more precise than previous fourth-order model due to the consideration of higher-order inertia integrals of the spacecraft. Some interesting conclusions about the gravity gradient torque model are reached. Then a numerical simulation is carried out to verify our model. In the numerical simulation, a special spacecraft consisted of 36 point masses connected by rigid massless rods is considered. We assume that the asteroid is in a uniform rotation around its maximum-moment principal axis, and the spacecraft is on the stationary orbit in the equatorial plane. Simulation results show that the motion of previous fourth-order model is quite different from the exact motion, while our full fourth-order model fits the exact motion very well. And our model is precise enough for practical applications.

**INTRODUCTION**

Studies on asteroids could provide answers to fundamental questions concerning the past of our Solar System. Over the past decades, interest in spacecraft missions to asteroids has increased. The spacecraft can make high-resolution observations and bring back samples, providing more detailed information than ground-based observations. Several missions have been developed with big success, such as NASA's Near Earth Asteroid Rendezvous (NEAR) mission to asteroid Eros and the JAXA (Japanese) mission Hayabusa to asteroid Itokawa. And several other missions are currently under development.

One of the key elements in designing such a mission is the analysis of dynamical behavior of the spacecraft around asteroids. An asteroid's irregular shape, non-spherical mass distribution and its rotational sate make the dynamics of the spacecraft quite complex and different from that around a large planet such as the Earth. Therefore, it is necessary to investigate the dynamics of the spacecraft around asteroids in details. The orbital dynamics of spacecraft around asteroids have been studied in many papers (for a recent review see Reference 1), while the attitude dynamics of spacecraft around asteroids have been studied by Kumar, Riverin and Misra (see References 1 and 2). The gravity gradient (GG) torque is the main perturbation of the attitude motion of the spacecraft. In this paper, we focus on the GG torque of the spacecraft in the non-central gravity field of the asteroid.

The GG torque of spacecraft about non-spherical bodies such as the Earth has been studied in several works (see References 3, 4 and 5). Their results showed that the main term of the GG torque was contributed by the central component of the gravity field of the Earth. The Earth's oblateness makes a contribution to the GG torque which is approximately 5 orders of magnitude less than the main term on the geosynchronous orbit. This is the reason why the oblateness of the Earth is not taken into consideration in the attitude dynamics of spacecraft around the Earth in theoretical studies and practical applications. However, the effects of the non-central gravity field of the asteroids on the attitude motion of the spacecraft can be significant and should be taken into consideration (see References 1 and 2).

In these previous studies on the GG torque in a non-central gravity field, inertia integrals of the spacecraft up to the second-order were considered. However, the third and fourth-order inertia integrals of the spacecraft, which have more significant effects on the GG torque than the non-central component of the gravity field, were not considered. As a result, only the second-order terms and parts of the fourth-order terms were included in the GG torque, with the third-order terms that were more significant and other fourth-order terms neglected. Thus, the previous model of the GG torque can be improved by taking into consideration of the spacecraft's higher-order inertia integrals.

In this paper, by taking into consideration of the spacecraft's inertia integrals up to the fourth-order, a full fourth-order model of the GG torque of spacecraft around asteroids is established. The gravity field of the asteroid is considered to be a 2nd degree and order-gravity field with harmonic coefficients $C_{20}$ and $C_{22}$. The fourth-order gravitational potential of the spacecraft is derived based on Taylor expansion. Then the expression of GG torque in terms of gravitational potential derivatives is derived. By using the formulation of the fourth-order gravitational potential derived above, explicit formulations of the full fourth-order GG torque are obtained. Based on the explicit formulations, interesting conclusions about the GG torque are reached. Then a numerical simulation, in which a special spacecraft consisted of 36 point masses is considered, is carried out to verify our model.

**STATEMENT OF THE PROBLEM**

As described in Figure 1, consider a rigid spacecraft $B$ moving around the asteroid $P$. The inertial reference frame is given by $S_i=\{e_1, e_2, e_3\}$ with $O_i$ as its origin. The body-fixed reference frames of the asteroid and the spacecraft are given by $S_P=\{u, v, w\}$ and $S_B=\{i, j, k\}$ with $O$ and $C$ as their origins respectively. The origin of the reference frame $S_P$ is at the mass center of the asteroid, and the coordinate axes are chosen to be aligned along the asteroid's principal moments of inertia. The principal moments of inertia of the asteroid are assumed to satisfy the following inequation

$$I_{P,zz} > I_{P,yy} > I_{P,xx} \tag{1}$$

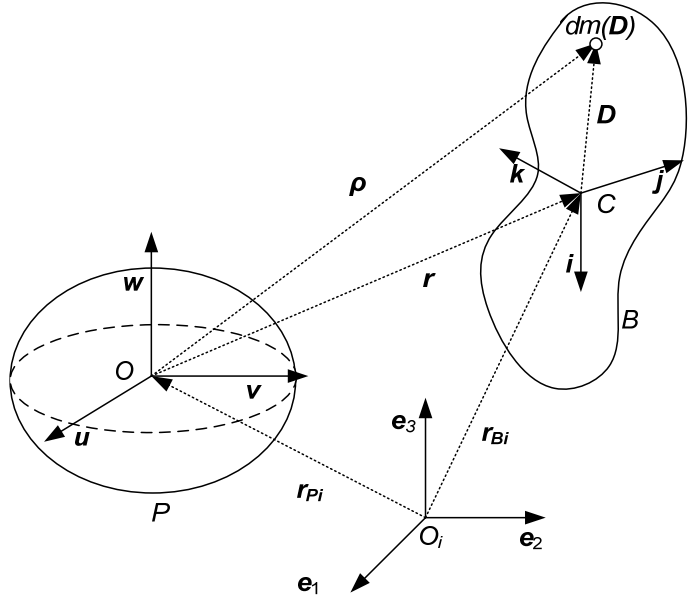

**Figure 1. A Rigid Spacecraft *B* around the Asteroid *P*.**

Then the 2nd degree and order-gravity field of the asteroid can be represented by the harmonic coefficients $C_{20}$ and $C_{22}$ with other harmonic coefficients vanished. The harmonic coefficients $C_{20}$ and $C_{22}$ are defined by

$$C_{20} = -\frac{1}{2Ma_e^2}\left(2I_{P,zz} - I_{P,xx} - I_{P,yy}\right) < 0 \tag{2}$$

$$C_{22} = \frac{1}{4Ma_e^2}\left(I_{P,yy} - I_{P,xx}\right) > 0 \tag{3}$$

where $M$ and $a_e$ are the mass and mean equatorial radius of the asteroid respectively. Also the reference frame $S_B$ is attached to the spacecraft's mass center and coincides with its principal axes reference frame.

The attitude matrices of $S_P$ and $S_B$ with respect to the inertial reference frame $S_i$ are denoted by $A_P$ and $A_B$ respectively

$$A_P = [\boldsymbol{u}_i, \boldsymbol{v}_i, \boldsymbol{w}_i] = \begin{bmatrix} u_i^x & v_i^x & w_i^x \\ u_i^y & v_i^y & w_i^y \\ u_i^z & v_i^z & w_i^z \end{bmatrix} \in SO(3), \quad A_B = [\boldsymbol{i}_i, \boldsymbol{j}_i, \boldsymbol{k}_i] = \begin{bmatrix} i_i^x & j_i^x & k_i^x \\ i_i^y & j_i^y & k_i^y \\ i_i^z & j_i^z & k_i^z \end{bmatrix} \in SO(3) \tag{4}$$

where $\boldsymbol{u}_i, \boldsymbol{v}_i, \boldsymbol{w}_i, \boldsymbol{i}_i, \boldsymbol{j}_i$ and $\boldsymbol{k}_i$ are coordinates of the unit vectors $\boldsymbol{u}, \boldsymbol{v}, \boldsymbol{w}, \boldsymbol{i}, \boldsymbol{j}$ and $\boldsymbol{k}$ in the inertial reference frame $S_i$ respectively. $SO(3)$ is the 3-dimensional special orthogonal group. The matrices $A_P$ and $A_B$ are also the coordinate transformation matrices from the corresponding body-fixed reference frame to the inertial reference frame $S_i$. The relative attitude matrix of the spacecraft with respect to the asteroid is given by

$$\boldsymbol{C} = A_P^T A_B \tag{5}$$

The attitude matrices $A_P$, $A_B$ and $C$ can be also written as follows

$$A_P = [\boldsymbol{\alpha}_P, \boldsymbol{\beta}_P, \boldsymbol{\gamma}_P]^T = \begin{bmatrix} \alpha_P^x & \alpha_P^y & \alpha_P^z \\ \beta_P^x & \beta_P^y & \beta_P^z \\ \gamma_P^x & \gamma_P^y & \gamma_P^z \end{bmatrix}, \quad A_B = [\boldsymbol{\alpha}_B, \boldsymbol{\beta}_B, \boldsymbol{\gamma}_B]^T = \begin{bmatrix} \alpha_B^x & \alpha_B^y & \alpha_B^z \\ \beta_B^x & \beta_B^y & \beta_B^z \\ \gamma_B^x & \gamma_B^y & \gamma_B^z \end{bmatrix} \tag{6}$$

$$C = [\boldsymbol{\alpha}, \boldsymbol{\beta}, \boldsymbol{\gamma}]^T = \begin{bmatrix} \alpha^x & \alpha^y & \alpha^z \\ \beta^x & \beta^y & \beta^z \\ \gamma^x & \gamma^y & \gamma^z \end{bmatrix} \tag{7}$$

where $\boldsymbol{\alpha}$, $\boldsymbol{\beta}$ and $\boldsymbol{\gamma}$ are coordinates of the unit vectors $\boldsymbol{u}$, $\boldsymbol{v}$ and $\boldsymbol{w}$ in the spacecraft's body-fixed reference frame $S_B$. And the matrix $C$ is the coordinate transformation matrix from the frame $S_B$ to the asteroid's body-fixed reference frame $S_P$.

$\boldsymbol{r}_{Pi}$ and $\boldsymbol{r}_{Bi}$ are radius vectors of the asteroid's mass center $O$ and the spacecraft's mass center $C$ with respect to $O_i$ expressed in the inertial reference frame $S_i$ respectively. Then the radius vector of the spacecraft's mass center $C$ with respect to the asteroid's mass center $O$ expressed in the asteroid's body-fixed reference frame $S_P$, denoted by $\boldsymbol{r}$, can be calculated by

$$\boldsymbol{r} = A_P^T (\boldsymbol{r}_{Bi} - \boldsymbol{r}_{Pi}) \tag{8}$$

$\boldsymbol{D}$ is the radius vector of the mass element $dm(\boldsymbol{D})$ of the spacecraft with respect to the spacecraft's mass center $C$ expressed in the spacecraft's body-fixed reference frame $S_B$. Then the radius vector of the mass element $dm(\boldsymbol{D})$ with respect to the asteroid's mass center $O$ expressed in the asteroid's body-fixed reference frame $S_P$, denoted by $\boldsymbol{\rho}$, is given by

$$\boldsymbol{\rho} = \boldsymbol{r} + C\boldsymbol{D} \tag{9}$$

According to Equation (5), vectors $\boldsymbol{\alpha}$, $\boldsymbol{\beta}$ and $\boldsymbol{\gamma}$ can be written in terms of $A_P$ and $A_B$ as follows

$$\boldsymbol{\alpha} = \alpha_P^x \boldsymbol{\alpha}_B + \beta_P^x \boldsymbol{\beta}_B + \gamma_P^x \boldsymbol{\gamma}_B \tag{10}$$

$$\boldsymbol{\beta} = \alpha_P^y \boldsymbol{\alpha}_B + \beta_P^y \boldsymbol{\beta}_B + \gamma_P^y \boldsymbol{\gamma}_B \tag{11}$$

$$\boldsymbol{\gamma} = \alpha_P^z \boldsymbol{\alpha}_B + \beta_P^z \boldsymbol{\beta}_B + \gamma_P^z \boldsymbol{\gamma}_B \tag{12}$$

As described in Figure 2, the gravitational potential of a unit mass point particle in the gravity field of the asteroid is given by

$$V_{UMP} = -\frac{\mu}{r} - \frac{\mu}{r^3}\left[\tau_0\left(1 - \frac{3}{2}\cos^2\delta\right) + 3\tau_2 \cos^2\delta \cos 2\lambda\right] \tag{13}$$

where $\mu = GM$, $G$ is the Gravitational Constant, $\tau_0 = a_e^2 C_{20}$, $\tau_2 = a_e^2 C_{22}$, $r$ is the distance of the particle from the mass center of the asteroid, $\lambda$ and $\delta$ are longitude and latitude of the particle respectively. The longitude $\lambda$ is measured counterclockwise from the $\boldsymbol{u}$-axis in the $\boldsymbol{u}$-$\boldsymbol{v}$ plane, and the latitude $\delta$ is measured from the $\boldsymbol{u}$-$\boldsymbol{v}$ plane towards the $\boldsymbol{w}$-axis. The Equation (13) can be also written in terms of $x$, $y$ and $z$, the three components of the vector $\boldsymbol{r}$, as follows

$$V_{UMP} = -\frac{\mu}{r}\left[1 + \frac{\tau_0}{r^2}\left(\frac{3}{2}\frac{z^2}{r^2} - \frac{1}{2}\right) + \frac{3\tau_2}{r^2}\frac{x^2 - y^2}{r^2}\right] \tag{14}$$

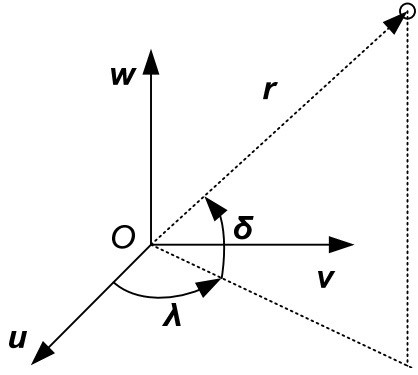

**Figure 2. A Unit Mass Point Particle in the Gravity Field of the Asteroid.**

The gravitational force of the unit mass point particle can be calculated by first order partial derivatives of the gravitational potential $V_{UMP}$ as follows

$$g^x = -\frac{\partial V_{UMP}}{\partial x} = -\frac{\mu x}{r^3} - \frac{\mu \tau_0 x}{r^5} + \frac{5\mu\tau_0 x\left(x^2 + y^2 - 2z^2\right)}{2r^7} + \frac{6\mu\tau_2 x}{r^5} - \frac{15\mu\tau_2 x\left(x^2 - y^2\right)}{r^7} \quad (15)$$

$$g^y = -\frac{\partial V_{UMP}}{\partial y} = -\frac{\mu y}{r^3} - \frac{\mu \tau_0 y}{r^5} + \frac{5\mu\tau_0 y\left(x^2 + y^2 - 2z^2\right)}{2r^7} - \frac{6\mu\tau_2 y}{r^5} - \frac{15\mu\tau_2 y\left(x^2 - y^2\right)}{r^7} \quad (16)$$

$$g^z = -\frac{\partial V_{UMP}}{\partial z} = -\frac{\mu z}{r^3} + \frac{2\mu\tau_0 z}{r^5} + \frac{5\mu\tau_0 z\left(x^2 + y^2 - 2z^2\right)}{2r^7} - \frac{15\mu\tau_2 z\left(x^2 - y^2\right)}{r^7} \quad (17)$$

**MUTUAL GRAVITATIONAL POTENTIAL**

According to Equations (9) and (14), the gravitational potential of the mass element $dm(\mathbf{D})$ of the spacecraft is given by

$$dV = -\frac{\mu dm(\mathbf{D})}{\rho}\left[1 + \frac{\tau_0}{\rho^2}\left(\frac{3}{2}\frac{\left(\rho^z\right)^2}{\rho^2} - \frac{1}{2}\right) + 3\tau_2 \frac{\left(\rho^x\right)^2 - \left(\rho^y\right)^2}{\rho^4}\right] \quad (18)$$

Using Equation (9) and the matrix $\mathbf{C}^T$ as the coordinate transformation matrix from the body-fixed frame $S_P$ to the body-fixed reference frame $S_B$, we have

$$\mathbf{P} = \mathbf{C}^T \boldsymbol{\rho} = \mathbf{C}^T \mathbf{r} + \mathbf{D} = \mathbf{R} + \mathbf{D} \quad (19)$$

where $\mathbf{P}$ and $\mathbf{R}$ are components of vectors $\boldsymbol{\rho}$ and $\mathbf{r}$ in the spacecraft's body-fixed frame $S_B$ respectively. Then $\rho$, $\rho^x$, $\rho^y$ and $\rho^z$ can be written in terms of $\mathbf{R} + \mathbf{D}$ and matrix $\mathbf{C}$ as follows

$$\rho = |\mathbf{R} + \mathbf{D}|, \quad \rho^x = \boldsymbol{\alpha} \cdot (\mathbf{R} + \mathbf{D}), \quad \rho^y = \boldsymbol{\beta} \cdot (\mathbf{R} + \mathbf{D}), \quad \rho^z = \boldsymbol{\gamma} \cdot (\mathbf{R} + \mathbf{D}) \quad (20)$$

Therefore, the gravitational potential of the mass element $dm(\mathbf{D})$ can be written as

$$dV = -\frac{\mu dm(\mathbf{D})}{|\mathbf{R} + \mathbf{D}|}\left[1 + \frac{\tau_0}{|\mathbf{R} + \mathbf{D}|^2}\left(\frac{3}{2}\frac{\left(\boldsymbol{\gamma} \cdot (\mathbf{R} + \mathbf{D})\right)^2}{|\mathbf{R} + \mathbf{D}|^2} - \frac{1}{2}\right) + 3\tau_2 \frac{\left(\boldsymbol{\alpha} \cdot (\mathbf{R} + \mathbf{D})\right)^2 - \left(\boldsymbol{\beta} \cdot (\mathbf{R} + \mathbf{D})\right)^2}{|\mathbf{R} + \mathbf{D}|^4}\right] \quad (21)$$

The gravitational potential of the spacecraft $B$ is formulated by the following integration over $B$

$$V(\boldsymbol{R}, \boldsymbol{\alpha}, \boldsymbol{\beta}, \boldsymbol{\gamma}) = \int_B dV(\boldsymbol{R}, \boldsymbol{\alpha}, \boldsymbol{\beta}, \boldsymbol{\gamma}, \boldsymbol{D}) \tag{22}$$

According to $|\boldsymbol{R} + \boldsymbol{D}|^2 = R^2 + D^2 + 2\boldsymbol{R} \cdot \boldsymbol{D}$, $|\boldsymbol{R} + \boldsymbol{D}|$ can be written as

$$|\boldsymbol{R} + \boldsymbol{D}| = R\left(1 + \frac{2\bar{\boldsymbol{R}} \cdot \boldsymbol{D}}{R} + \frac{D^2}{R^2}\right)^{\frac{1}{2}} \tag{23}$$

where $\bar{\boldsymbol{R}}$ is the unit vector along the vector $\boldsymbol{R}$. By using Equation (23), the expressions about $|\boldsymbol{R} + \boldsymbol{D}|$ in Equation (21) can be written in the form of series through Taylor expansion and truncation on the appropriate order

$$\frac{1}{|\boldsymbol{R} + \boldsymbol{D}|} = \frac{1}{R} - \frac{\bar{\boldsymbol{R}} \cdot \boldsymbol{D}}{R^2} + \left[\frac{3}{2}\frac{(\bar{\boldsymbol{R}} \cdot \boldsymbol{D})^2}{R^3} - \frac{1}{2}\frac{D^2}{R^3}\right] + \left[\frac{3}{2}\frac{D^2(\bar{\boldsymbol{R}} \cdot \boldsymbol{D})}{R^4} - \frac{5}{2}\frac{(\bar{\boldsymbol{R}} \cdot \boldsymbol{D})^3}{R^4}\right]$$
$$+ \left[\frac{3}{8}\frac{D^4}{R^5} - \frac{15}{4}\frac{D^2(\bar{\boldsymbol{R}} \cdot \boldsymbol{D})^2}{R^5} + \frac{35}{8}\frac{(\bar{\boldsymbol{R}} \cdot \boldsymbol{D})^4}{R^5}\right] + O(R^{-6}) \tag{24}$$

$$\frac{1}{|\boldsymbol{R} + \boldsymbol{D}|^3} = \frac{1}{R^3} - 3\frac{\bar{\boldsymbol{R}} \cdot \boldsymbol{D}}{R^4} + \left[\frac{15}{2}\frac{(\bar{\boldsymbol{R}} \cdot \boldsymbol{D})^2}{R^5} - \frac{3}{2}\frac{D^2}{R^5}\right] + O(R^{-6}) \tag{25}$$

$$\frac{(\boldsymbol{\alpha} \cdot (\boldsymbol{R} + \boldsymbol{D}))^2}{|\boldsymbol{R} + \boldsymbol{D}|^5} = \frac{(\boldsymbol{\alpha} \cdot \bar{\boldsymbol{R}})^2}{R^3} + \left(\frac{2(\boldsymbol{\alpha} \cdot \bar{\boldsymbol{R}})(\boldsymbol{\alpha} \cdot \boldsymbol{D})}{R^4} - 5\frac{(\boldsymbol{\alpha} \cdot \bar{\boldsymbol{R}})^2(\bar{\boldsymbol{R}} \cdot \boldsymbol{D})}{R^4}\right)$$
$$+ \left[\frac{(\boldsymbol{\alpha} \cdot \boldsymbol{D})^2}{R^5} - \frac{5}{2}\frac{(\boldsymbol{\alpha} \cdot \bar{\boldsymbol{R}})^2 D^2}{R^5} + \frac{35}{2}\frac{(\boldsymbol{\alpha} \cdot \bar{\boldsymbol{R}})^2(\bar{\boldsymbol{R}} \cdot \boldsymbol{D})^2}{R^5} - \frac{10(\boldsymbol{\alpha} \cdot \bar{\boldsymbol{R}})(\boldsymbol{\alpha} \cdot \boldsymbol{D})(\bar{\boldsymbol{R}} \cdot \boldsymbol{D})}{R^5}\right] + O(R^{-6}) \tag{26}$$

$$\frac{(\boldsymbol{\beta} \cdot (\boldsymbol{R} + \boldsymbol{D}))^2}{|\boldsymbol{R} + \boldsymbol{D}|^5} = \frac{(\boldsymbol{\beta} \cdot \bar{\boldsymbol{R}})^2}{R^3} + \left(\frac{2(\boldsymbol{\beta} \cdot \bar{\boldsymbol{R}})(\boldsymbol{\beta} \cdot \boldsymbol{D})}{R^4} - 5\frac{(\boldsymbol{\beta} \cdot \bar{\boldsymbol{R}})^2(\bar{\boldsymbol{R}} \cdot \boldsymbol{D})}{R^4}\right)$$
$$+ \left[\frac{(\boldsymbol{\beta} \cdot \boldsymbol{D})^2}{R^5} - \frac{5}{2}\frac{(\boldsymbol{\beta} \cdot \bar{\boldsymbol{R}})^2 D^2}{R^5} + \frac{35}{2}\frac{(\boldsymbol{\beta} \cdot \bar{\boldsymbol{R}})^2(\bar{\boldsymbol{R}} \cdot \boldsymbol{D})^2}{R^5} - \frac{10(\boldsymbol{\beta} \cdot \bar{\boldsymbol{R}})(\boldsymbol{\beta} \cdot \boldsymbol{D})(\bar{\boldsymbol{R}} \cdot \boldsymbol{D})}{R^5}\right] + O(R^{-6}) \tag{27}$$

$$\frac{(\boldsymbol{\gamma} \cdot (\boldsymbol{R} + \boldsymbol{D}))^2}{|\boldsymbol{R} + \boldsymbol{D}|^5} = \frac{(\boldsymbol{\gamma} \cdot \bar{\boldsymbol{R}})^2}{R^3} + \left(\frac{2(\boldsymbol{\gamma} \cdot \bar{\boldsymbol{R}})(\boldsymbol{\gamma} \cdot \boldsymbol{D})}{R^4} - 5\frac{(\boldsymbol{\gamma} \cdot \bar{\boldsymbol{R}})^2(\bar{\boldsymbol{R}} \cdot \boldsymbol{D})}{R^4}\right)$$
$$+ \left[\frac{(\boldsymbol{\gamma} \cdot \boldsymbol{D})^2}{R^5} - \frac{5}{2}\frac{(\boldsymbol{\gamma} \cdot \bar{\boldsymbol{R}})^2 D^2}{R^5} + \frac{35}{2}\frac{(\boldsymbol{\gamma} \cdot \bar{\boldsymbol{R}})^2(\bar{\boldsymbol{R}} \cdot \boldsymbol{D})^2}{R^5} - \frac{10(\boldsymbol{\gamma} \cdot \bar{\boldsymbol{R}})(\boldsymbol{\gamma} \cdot \boldsymbol{D})(\bar{\boldsymbol{R}} \cdot \boldsymbol{D})}{R^5}\right] + O(R^{-6}) \tag{28}$$

By using Equations (24)-(28), the leading terms of $dV$ up to the fourth-order can be written as

$$dV^{(0)} = -\frac{\mu dm(\boldsymbol{D})}{R} \tag{29}$$

$$dV^{(1)} = \mu \frac{\bar{R} \cdot D}{R^2} dm(D) \tag{30}$$

$$dV^{(2)} = -\frac{\mu dm(D)}{2R^3}\left[3(\bar{R} \cdot D)^2 - D^2 - \tau_0 + 3\tau_0(\gamma \cdot \bar{R})^2 + 6\tau_2(\alpha \cdot \bar{R})^2 - 6\tau_2(\beta \cdot \bar{R})^2\right] \tag{31}$$

$$dV^{(3)} = -\frac{\mu dm(D)}{2R^4}\{(\bar{R} \cdot D)\left[3D^2 + 3\tau_0 - 5(\bar{R} \cdot D)^2\right] + 6\tau_0(\gamma \cdot \bar{R})(\gamma \cdot D) - 15\tau_0(\gamma \cdot \bar{R})^2(\bar{R} \cdot D)$$
$$+12\tau_2(\alpha \cdot \bar{R})(\alpha \cdot D) - 30\tau_2(\alpha \cdot \bar{R})^2(\bar{R} \cdot D) - 12\tau_2(\beta \cdot \bar{R})(\beta \cdot D) + 30\tau_2(\beta \cdot \bar{R})^2(\bar{R} \cdot D)\} \tag{32}$$

$$dV^{(4)} = -\frac{\mu dm(D)}{8R^5}\{\left[3D^4 - 30D^2(\bar{R} \cdot D)^2 + 35(\bar{R} \cdot D)^4\right] + 2\tau_0\left[3D^2 - 15(\bar{R} \cdot D)^2\right]$$
$$+12\tau_0(\gamma \cdot D)^2 - 30\tau_0(\gamma \cdot \bar{R})^2 D^2 + 210\tau_0(\gamma \cdot \bar{R})^2(\bar{R} \cdot D)^2 - 120\tau_0(\gamma \cdot \bar{R})(\gamma \cdot D)(\bar{R} \cdot D)$$
$$+24\tau_2(\alpha \cdot D)^2 - 60\tau_2(\alpha \cdot \bar{R})^2 D^2 + 420\tau_2(\alpha \cdot \bar{R})^2(\bar{R} \cdot D)^2 - 240\tau_2(\alpha \cdot \bar{R})(\alpha \cdot D)(\bar{R} \cdot D)$$
$$-24\tau_2(\beta \cdot D)^2 + 60\tau_2(\beta \cdot \bar{R})^2 D^2 - 420\tau_2(\beta \cdot \bar{R})^2(\bar{R} \cdot D)^2 + 240\tau_2(\beta \cdot \bar{R})(\beta \cdot D)(\bar{R} \cdot D)\} \tag{33}$$

Substitution of Equations (29)-(33) into Equation (22) gives the leading terms of the gravitational potential $V$, namely $V^{(0)}$ to $V^{(4)}$. The zeroth-order gravitational potential $V^{(0)}$ is given by

$$V^{(0)} = \int_B -\frac{\mu dm(D)}{R} = -\frac{\mu m}{R} \tag{34}$$

where $m$ is the mass of the spacecraft. Since the origin of the frame $S_B$ coincides with the mass center of the spacecraft, the first-order gravitational potential $V^{(1)}$ is vanished.

$$V^{(1)} = \mu \frac{\bar{R}}{R^2} \cdot \int_B D dm(D) = 0 \tag{35}$$

And the inertia integrals of the spacecraft $B$ are defined by

$$J_{\underbrace{x...x}_{p-times}\underbrace{y...y}_{q-times}\underbrace{z...z}_{r-times}} = \int_B (D^x)^p (D^y)^q (D^z)^r dm(D) \tag{36}$$

The moments of inertia are defined by $I_{xx} = J_{yy} + J_{zz}$, $I_{yy} = J_{xx} + J_{zz}$ and $I_{zz} = J_{xx} + J_{yy}$. The frame $S_B$ is the spacecraft's principal axes reference frame, thus the product moments of inertia are all eliminated. We express $\bar{R} \cdot D$, $D^2$, $\alpha \cdot D$, $\beta \cdot D$ and $\gamma \cdot D$ in terms of components in the body-fixed frame $S_B$ as follows

$$\bar{R} \cdot D = \bar{R}^x D^x + \bar{R}^y D^y + \bar{R}^z D^z, \quad D^2 = (D^x)^2 + (D^y)^2 + (D^z)^2 \tag{37}$$

$$\alpha \cdot D = \alpha^x D^x + \alpha^y D^y + \alpha^z D^z, \quad \beta \cdot D = \beta^x D^x + \beta^y D^y + \beta^z D^z \tag{38}$$

$$\gamma \cdot D = \gamma^x D^x + \gamma^y D^y + \gamma^z D^z \tag{39}$$

Then using inertia integrals defined above, we can get the second-order gravitational potential $V^{(2)}$ and the third-order gravitational potential $V^{(3)}$ as follows

$$V^{(2)}(\mathbf{R}, \boldsymbol{\alpha}, \boldsymbol{\beta}, \boldsymbol{\gamma}) = -\frac{\mu}{2R^3}\Big[\big(1-3(\bar{R}^x)^2\big)I_{xx} + \big(1-3(\bar{R}^y)^2\big)I_{yy} + \big(1-3(\bar{R}^z)^2\big)I_{zz} - m\tau_0$$
$$+3m\tau_0(\boldsymbol{\gamma}\cdot\bar{\mathbf{R}})^2 + 6m\tau_2(\boldsymbol{\alpha}\cdot\bar{\mathbf{R}})^2 - 6m\tau_2(\boldsymbol{\beta}\cdot\bar{\mathbf{R}})^2\Big] \quad (40)$$

$$V^{(3)}(\mathbf{R}, \boldsymbol{\alpha}, \boldsymbol{\beta}, \boldsymbol{\gamma}) = \frac{\mu}{2R^4}\Big[\bar{R}^x\big(5(\bar{R}^x)^2-3\big)J_{xxx} + \bar{R}^y\big(5(\bar{R}^y)^2-3\big)J_{yyy} + \bar{R}^z\big(5(\bar{R}^z)^2-3\big)J_{zzz}$$
$$+3\bar{R}^y\big(5(\bar{R}^x)^2-1\big)J_{xxy} + 3\bar{R}^x\big(5(\bar{R}^y)^2-1\big)J_{xyy} + 3\bar{R}^z\big(5(\bar{R}^x)^2-1\big)J_{xxz} + 3\bar{R}^x\big(5(\bar{R}^z)^2-1\big)J_{xzz}$$
$$+3\bar{R}^z\big(5(\bar{R}^y)^2-1\big)J_{yyz} + 3\bar{R}^y\big(5(\bar{R}^z)^2-1\big)J_{yzz} + 30\bar{R}^x\bar{R}^y\bar{R}^z J_{xyz}\Big] \quad (41)$$

The fourth-order potential $V^{(4)}$ can be written into three parts as follows

$$V^{(4),C}(\mathbf{R}, \boldsymbol{\alpha}, \boldsymbol{\beta}, \boldsymbol{\gamma}) = -\frac{\mu}{8R^5}\Big[\big(35(\bar{R}^x)^4 - 30(\bar{R}^x)^2 + 3\big)J_{xxxx} + \big(35(\bar{R}^y)^4 - 30(\bar{R}^y)^2 + 3\big)J_{yyyy}$$
$$+20\bar{R}^x\bar{R}^z\big(7(\bar{R}^z)^2-3\big)J_{xzzz} + 20\bar{R}^y\bar{R}^z\big(7(\bar{R}^y)^2-3\big)J_{yyyz} + 20\bar{R}^y\bar{R}^z\big(7(\bar{R}^z)^2-3\big)J_{yzzz}$$
$$+\big(35(\bar{R}^z)^4 - 30(\bar{R}^z)^2 + 3\big)J_{zzzz} + 20\bar{R}^x\bar{R}^y\big(7(\bar{R}^y)^2-3\big)J_{xyyy} + 20\bar{R}^x\bar{R}^z\big(7(\bar{R}^x)^2-3\big)J_{xxxz}$$
$$+6\big(35(\bar{R}^x)^2(\bar{R}^y)^2 - 5(\bar{R}^x)^2 - 5(\bar{R}^y)^2 + 1\big)J_{xxyy} + 20\bar{R}^x\bar{R}^y\big(7(\bar{R}^x)^2-3\big)J_{xxxy}$$
$$+6\big(35(\bar{R}^x)^2(\bar{R}^z)^2 - 5(\bar{R}^x)^2 - 5(\bar{R}^z)^2 + 1\big)J_{xxzz} + 60\bar{R}^y\bar{R}^z\big(7(\bar{R}^x)^2-1\big)J_{xxyz}$$
$$+6\big(35(\bar{R}^y)^2(\bar{R}^z)^2 - 5(\bar{R}^y)^2 - 5(\bar{R}^z)^2 + 1\big)J_{yyzz} + 60\bar{R}^x\bar{R}^z\big(7(\bar{R}^y)^2-1\big)J_{xyyz}$$
$$+60\bar{R}^x\bar{R}^y\big(7(\bar{R}^z)^2-1\big)J_{xyzz}\Big] \quad (42)$$

$$V^{(4),C_{20}}(\mathbf{R}, \boldsymbol{\alpha}, \boldsymbol{\beta}, \boldsymbol{\gamma}) = -\frac{\mu}{8R^5}\Big[6\tau_0\big(15(\boldsymbol{\gamma}\cdot\bar{\mathbf{R}})^2 - 2\big)(I_{xx} + I_{yy} + I_{zz})$$
$$+30\tau_0\big(1-7(\boldsymbol{\gamma}\cdot\bar{\mathbf{R}})^2\big)\big((\bar{R}^x)^2 I_{xx} + (\bar{R}^y)^2 I_{yy} + (\bar{R}^z)^2 I_{zz}\big) + 6\tau_0\gamma^x\big(\gamma^x - 10(\boldsymbol{\gamma}\cdot\bar{\mathbf{R}})\bar{R}^x\big)\big(-I_{xx} + I_{yy} + I_{zz}\big)$$
$$+6\tau_0\gamma^z\big(\gamma^z - 10(\boldsymbol{\gamma}\cdot\bar{\mathbf{R}})\bar{R}^z\big)\big(I_{xx} + I_{yy} - I_{zz}\big) + 6\tau_0\gamma^y\big(\gamma^y - 10(\boldsymbol{\gamma}\cdot\bar{\mathbf{R}})\bar{R}^y\big)\big(I_{xx} - I_{yy} + I_{zz}\big)\Big] \quad (43)$$

$$V^{(4),C_{22}}(\mathbf{R}, \boldsymbol{\alpha}, \boldsymbol{\beta}, \boldsymbol{\gamma}) = -\frac{\mu}{8R^5}\Big[180\tau_2(\boldsymbol{\alpha}\cdot\bar{\mathbf{R}})^2(I_{xx} + I_{yy} + I_{zz})$$
$$-420\tau_2(\boldsymbol{\alpha}\cdot\bar{\mathbf{R}})^2\big((\bar{R}^x)^2 I_{xx} + (\bar{R}^y)^2 I_{yy} + (\bar{R}^z)^2 I_{zz}\big) + 12\tau_2\alpha^x\big(\alpha^x - 10(\boldsymbol{\alpha}\cdot\bar{\mathbf{R}})\bar{R}^x\big)\big(-I_{xx} + I_{yy} + I_{zz}\big)$$
$$+12\tau_2\alpha^y\big(\alpha^y - 10(\boldsymbol{\alpha}\cdot\bar{\mathbf{R}})\bar{R}^y\big)\big(I_{xx} - I_{yy} + I_{zz}\big) + 12\tau_2\alpha^z\big(\alpha^z - 10(\boldsymbol{\alpha}\cdot\bar{\mathbf{R}})\bar{R}^z\big)\big(I_{xx} + I_{yy} - I_{zz}\big)$$
$$-180\tau_2(\boldsymbol{\beta}\cdot\bar{\mathbf{R}})^2(I_{xx} + I_{yy} + I_{zz}) + 420\tau_2(\boldsymbol{\beta}\cdot\bar{\mathbf{R}})^2\big((\bar{R}^x)^2 I_{xx} + (\bar{R}^y)^2 I_{yy} + (\bar{R}^z)^2 I_{zz}\big)$$
$$-12\tau_2\beta^x\big(\beta^x - 10(\boldsymbol{\beta}\cdot\bar{\mathbf{R}})\bar{R}^x\big)\big(-I_{xx} + I_{yy} + I_{zz}\big) - 12\tau_2\beta^y\big(\beta^y - 10(\boldsymbol{\beta}\cdot\bar{\mathbf{R}})\bar{R}^y\big)\big(I_{xx} - I_{yy} + I_{zz}\big)$$
$$-12\tau_2\beta^z\big(\beta^z - 10(\boldsymbol{\beta}\cdot\bar{\mathbf{R}})\bar{R}^z\big)\big(I_{xx} + I_{yy} - I_{zz}\big)\Big] \quad (44)$$

$V^{(4),C}$ is the gravitational potential due to the interaction between the central component of *P*'s gravity field and the spacecraft's fourth-order inertia integrals; $V^{(4),C_{20}}$ is the gravitational potential of the interaction between the second degree and zeroth order component of *P*'s gravity field and the spacecraft's second-order inertia integrals; $V^{(4),C_{22}}$ is due to the interaction between the second degree and second order component of *P*'s gravity field and the spacecraft's second-order inertia integrals.

The fourth-order approximate gravitational potential $\tilde{V}$ is the sum of $V^{(0)}$, $V^{(2)}$, $V^{(3)}$, $V^{(4),C}$, $V^{(4),C_{20}}$ and $V^{(4),C_{22}}$

$$\tilde{V}(\boldsymbol{R}, \boldsymbol{\alpha}, \boldsymbol{\beta}, \boldsymbol{\gamma}) = V^{(0)} + V^{(2)} + V^{(3)} + V^{(4),C} + V^{(4),C_{20}} + V^{(4),C_{22}} \tag{45}$$

## GRAVITY GRADIENT TORQUE

After the formulation of the mutual gravitational potential between the asteroid and the spacecraft obtained, the explicit formulations of the GG torque acting on the spacecraft can be derived through the gravitational potential derivatives. The gravitational potential $\tilde{V}$ is a function of the inertial positions and attitudes of the asteroid and the spacecraft $\boldsymbol{r}_{Pi}$, $\boldsymbol{r}_{Bi}$, $\boldsymbol{A}_P$ and $\boldsymbol{A}_B$. However, according to Equations (34), (40)-(45) we notice that gravitational potential $\tilde{V}$ can be determined by the relative position and attitude of the spacecraft with respect to the asteroid. That is to say

$$\tilde{V} = \tilde{V}(\boldsymbol{r}_{Pi}, \boldsymbol{r}_{Bi}, \boldsymbol{A}_P, \boldsymbol{A}_B) = \tilde{V}(\boldsymbol{R}, \boldsymbol{C}) \tag{46}$$

And we have several relations between these two sets of variables of position and attitude, as shown by Equations (10)-(12) and the following equation

$$\boldsymbol{R} = \boldsymbol{A}_B^T (\boldsymbol{r}_{Bi} - \boldsymbol{r}_{Pi}) \tag{47}$$

When the mutual gravitational potential $\tilde{V}$ is considered as a function of $\boldsymbol{r}_{Pi}$, $\boldsymbol{r}_{Bi}$, $\boldsymbol{A}_P$ and $\boldsymbol{A}_B$, the fourth-order approximate GG torque acting on the spacecraft *B* expressed in the body-fixed frame $S_B$, denoted by $\tilde{\boldsymbol{T}}_B$, can be calculated by (see Reference 6)

$$\tilde{\boldsymbol{T}}_B = \boldsymbol{\alpha}_B \times \frac{\partial \tilde{V}(\boldsymbol{r}_{Pi}, \boldsymbol{r}_{Bi}, \boldsymbol{A}_P, \boldsymbol{A}_B)}{\partial \boldsymbol{\alpha}_B} + \boldsymbol{\beta}_B \times \frac{\partial \tilde{V}(\boldsymbol{r}_{Pi}, \boldsymbol{r}_{Bi}, \boldsymbol{A}_P, \boldsymbol{A}_B)}{\partial \boldsymbol{\beta}_B} + \boldsymbol{\gamma}_B \times \frac{\partial \tilde{V}(\boldsymbol{r}_{Pi}, \boldsymbol{r}_{Bi}, \boldsymbol{A}_P, \boldsymbol{A}_B)}{\partial \boldsymbol{\gamma}_B} \tag{48}$$

According to Equations (10)-(12) and (47), we obtain following equations using the chain rule

$$\frac{\partial \tilde{V}(\boldsymbol{r}_{Pi}, \boldsymbol{r}_{Bi}, \boldsymbol{A}_P, \boldsymbol{A}_B)}{\partial \boldsymbol{\alpha}_B} = \left(\frac{\partial \boldsymbol{R}}{\partial \boldsymbol{\alpha}_B}\right)^T \frac{\partial \tilde{V}(\boldsymbol{R}, \boldsymbol{C})}{\partial \boldsymbol{R}} + \left(\frac{\partial \boldsymbol{\alpha}}{\partial \boldsymbol{\alpha}_B}\right)^T \frac{\partial \tilde{V}(\boldsymbol{R}, \boldsymbol{C})}{\partial \boldsymbol{\alpha}} + \left(\frac{\partial \boldsymbol{\beta}}{\partial \boldsymbol{\alpha}_B}\right)^T \frac{\partial \tilde{V}(\boldsymbol{R}, \boldsymbol{C})}{\partial \boldsymbol{\beta}}$$
$$+ \left(\frac{\partial \boldsymbol{\gamma}}{\partial \boldsymbol{\alpha}_B}\right)^T \frac{\partial \tilde{V}(\boldsymbol{R}, \boldsymbol{C})}{\partial \boldsymbol{\gamma}} = (\boldsymbol{r}_{Bi} - \boldsymbol{r}_{Pi})^x \frac{\partial \tilde{V}(\boldsymbol{R}, \boldsymbol{C})}{\partial \boldsymbol{R}} + \alpha_P^x \frac{\partial \tilde{V}(\boldsymbol{R}, \boldsymbol{C})}{\partial \boldsymbol{\alpha}} + \alpha_P^y \frac{\partial \tilde{V}(\boldsymbol{R}, \boldsymbol{C})}{\partial \boldsymbol{\beta}} + \alpha_P^z \frac{\partial \tilde{V}(\boldsymbol{R}, \boldsymbol{C})}{\partial \boldsymbol{\gamma}} \tag{49}$$

$$\frac{\partial \tilde{V}(\boldsymbol{r}_{Pi}, \boldsymbol{r}_{Bi}, \boldsymbol{A}_P, \boldsymbol{A}_B)}{\partial \boldsymbol{\beta}_B} = \left(\frac{\partial \boldsymbol{R}}{\partial \boldsymbol{\beta}_B}\right)^T \frac{\partial \tilde{V}(\boldsymbol{R}, \boldsymbol{C})}{\partial \boldsymbol{R}} + \left(\frac{\partial \boldsymbol{\alpha}}{\partial \boldsymbol{\beta}_B}\right)^T \frac{\partial \tilde{V}(\boldsymbol{R}, \boldsymbol{C})}{\partial \boldsymbol{\alpha}} + \left(\frac{\partial \boldsymbol{\beta}}{\partial \boldsymbol{\beta}_B}\right)^T \frac{\partial \tilde{V}(\boldsymbol{R}, \boldsymbol{C})}{\partial \boldsymbol{\beta}}$$
$$+ \left(\frac{\partial \boldsymbol{\gamma}}{\partial \boldsymbol{\beta}_B}\right)^T \frac{\partial \tilde{V}(\boldsymbol{R}, \boldsymbol{C})}{\partial \boldsymbol{\gamma}} = (\boldsymbol{r}_{Bi} - \boldsymbol{r}_{Pi})^y \frac{\partial \tilde{V}(\boldsymbol{R}, \boldsymbol{C})}{\partial \boldsymbol{R}} + \beta_P^x \frac{\partial \tilde{V}(\boldsymbol{R}, \boldsymbol{C})}{\partial \boldsymbol{\alpha}} + \beta_P^y \frac{\partial \tilde{V}(\boldsymbol{R}, \boldsymbol{C})}{\partial \boldsymbol{\beta}} + \beta_P^z \frac{\partial \tilde{V}(\boldsymbol{R}, \boldsymbol{C})}{\partial \boldsymbol{\gamma}} \tag{50}$$

$$\frac{\partial \tilde{V}(r_{Pi}, r_{Bi}, A_P, A_B)}{\partial \gamma_B} = \left(\frac{\partial R}{\partial \gamma_B}\right)^T \frac{\partial \tilde{V}(R,C)}{\partial R} + \left(\frac{\partial \alpha}{\partial \gamma_B}\right)^T \frac{\partial \tilde{V}(R,C)}{\partial \alpha} + \left(\frac{\partial \beta}{\partial \gamma_B}\right)^T \frac{\partial \tilde{V}(R,C)}{\partial \beta}$$

$$+ \left(\frac{\partial \gamma}{\partial \gamma_B}\right)^T \frac{\partial \tilde{V}(R,C)}{\partial \gamma} = (r_{Bi} - r_{Pi})^z \frac{\partial \tilde{V}(R,C)}{\partial R} + \gamma_P^x \frac{\partial \tilde{V}(R,C)}{\partial \alpha} + \gamma_P^y \frac{\partial \tilde{V}(R,C)}{\partial \beta} + \gamma_P^z \frac{\partial \tilde{V}(R,C)}{\partial \gamma} \quad (51)$$

where $\partial a/\partial b = \begin{bmatrix} \partial a^x/\partial b & \partial a^y/\partial b & \partial a^z/\partial b \end{bmatrix}^T$ is the Jacobi matrix. Substitution of Equations (49)-(51) into Equation (48) gives

$$\tilde{T}_B = R \times \frac{\partial \tilde{V}(R,C)}{\partial R} + \alpha \times \frac{\partial \tilde{V}(R,C)}{\partial \alpha} + \beta \times \frac{\partial \tilde{V}(R,C)}{\partial \beta} + \gamma \times \frac{\partial \tilde{V}(R,C)}{\partial \gamma} \quad (52)$$

The explicit formulations of $\tilde{T}_B$ are obtained by using Equation (52) with the help of *Maple*. These formulations are given in the APPENDIX. It is found that every term in the formulations of the gravitational potential and torque contains a product of two mass distribution parameters, among which one is the asteroid's and the other is the spacecraft's. And the order of the term is sum of orders of the two mass distribution parameters. For the asteroid, the zeroth-order mass distribution parameter is the mass $M$; the second-order parameters are $C_{20}$ and $C_{22}$. For the spacecraft, the mass distribution parameters are inertia integrals.

The spacecraft's zeroth-order inertia integral, i.e. the spacecraft's mass, has no contribution to the GG torque, and the spacecraft's first-order inertia integrals are vanished. Then we can conclude that the harmonic coefficients of the asteroid's gravity field higher than second-order have no contribution to the fourth-order GG torque model. Therefore, the assumption of a 2nd degree and order-gravity field is precise enough for a fourth-order GG torque model. The coefficients $C_{20}$ and $C_{22}$ appear in the fourth-order terms of the GG torque along with the second-order inertia integrals of the spacecraft. These conclusions are verified by Equations (A.1)-(A.3) in APPENDIX.

The third and fourth-order inertia integrals of the spacecraft appear in the third and fourth-order terms of the GG torque respectively along with the asteroid's mass. In previous results (see References 1-5) the third and fourth-order inertia integrals of the spacecraft were not considered, thus only the second-order and some fourth-order terms of the GG torque were included with the third-order terms and parts of fourth-order terms neglected. Therefore, our full fourth-order GG torque model is more sound and precise than previous fourth-order model. This conclusion is confirmed by our numerical simulation in the next section.

## SIMULATION EXAMPLE

A numerical simulation is carried out to verify our GG torque model. We assume that the mass center of the asteroid is stationary in the inertial space, and the asteroid is in a uniform rotation around its maximum-moment principal axis, i.e. the *w*-axis. The spacecraft is assumed to be on a stationary orbit and the orbit motion is negligibly affected by the attitude motion, as described by Figure 3. According to orbital theory, a stationary orbit in inertial space corresponds to an equilibrium in the asteroid's body-fixed frame. There are two kinds of stationary orbits: those that lie on the *u*-axis, and those that lie on the *v*-axis. The stationary orbits lying on the *u*-axis are always unstable, while those lying on the *v*-axis are stable under the following condition

$$\left(\frac{\mu}{\omega_T^2}\right)^{\frac{2}{3}} + \tau_0 - 162\tau_2 > 0 \quad (53)$$

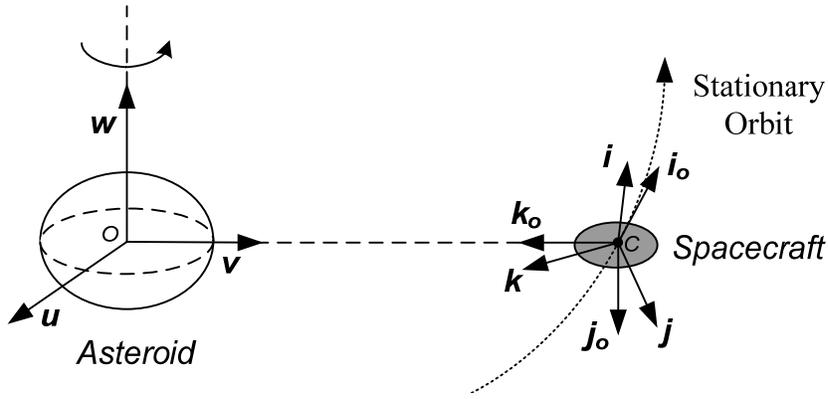

**Figure 3. The Spacecraft on a Stationary Orbit around the Asteroid.**

where $\omega_T$ is the angular velocity of the asteroid's uniform rotation. In the numerical simulation, we consider a stable stationary orbit, i.e. the spacecraft is located on the *v*-axis. The radius of the stationary orbit $R_S$ satisfies the following equation

$$R_S^5 - \frac{\mu}{\omega_T^2}\left(R_S^2 - \frac{3}{2}\tau_0 - 9\tau_2\right) = 0 \tag{54}$$

As described by Figure 3, the orbital reference frame is defined by $S_O = \{i_o, j_o, k_o\}$ with its origin coinciding with *C*, mass center of the spacecraft. And $k_o$ points towards the mass center of the asteroid, $j_o$ is in the opposite direction of the orbital angular momentum, and $i_o$ completes the orthogonal triad. The attitude of the spacecraft with respect to the orbital frame is defined in terms of roll, pitch and yaw angles. The sequence of rotation is: yaw $\psi$ around the *k*-axis, followed by pitch $\theta$ around the *j*-axis, and then roll $\phi$ around the *i*-axis. The sequence of rotation from the reference frame $S_P$ to frame $S_O$, then to frame $S_B$ can be described as follows

$$S_P \xrightarrow{R_x(\frac{\pi}{2})} \circ \xrightarrow{R_z(\pi)} S_O \xrightarrow{R_z(\psi)} \circ \xrightarrow{R_y(\theta)} \circ \xrightarrow{R_x(\phi)} S_B$$

The coordinate transformation matrix from the asteroid's body-fixed frame $S_P$ to the orbital reference frame $S_O$ is given by

$$\mathbf{L}_{OP} = \mathbf{L}_z(\pi)\mathbf{L}_x(\frac{\pi}{2}) = \begin{bmatrix} -1 & 0 & 0 \\ 0 & -1 & 0 \\ 0 & 0 & 1 \end{bmatrix}\begin{bmatrix} 1 & 0 & 0 \\ 0 & 0 & 1 \\ 0 & -1 & 0 \end{bmatrix} = \begin{bmatrix} -1 & 0 & 0 \\ 0 & 0 & -1 \\ 0 & -1 & 0 \end{bmatrix} \tag{55}$$

And the coordinate transformation matrix from the orbital reference frame $S_O$ to the spacecraft's body-fixed frame $S_B$ is given by

$$\mathbf{L}_{BO} = \mathbf{L}_x(\phi)\mathbf{L}_y(\theta)\mathbf{L}_z(\psi) = \begin{bmatrix} 1 & 0 & 0 \\ 0 & \cos\phi & \sin\phi \\ 0 & -\sin\phi & \cos\phi \end{bmatrix}\begin{bmatrix} \cos\theta & 0 & -\sin\theta \\ 0 & 1 & 0 \\ \sin\theta & 0 & \cos\theta \end{bmatrix}\begin{bmatrix} \cos\psi & \sin\psi & 0 \\ -\sin\psi & \cos\psi & 0 \\ 0 & 0 & 1 \end{bmatrix} \tag{56}$$

We assume further that the GG torque is the only torque acting on the spacecraft. The equations of attitude motion are given as follows

$$I_{xx}\dot{\Omega}^x - (I_{yy} - I_{zz})\Omega^y\Omega^z = T_B^x$$
$$I_{yy}\dot{\Omega}^y - (I_{zz} - I_{xx})\Omega^x\Omega^z = T_B^y \qquad (57)$$
$$I_{zz}\dot{\Omega}^z - (I_{xx} - I_{yy})\Omega^x\Omega^y = T_B^z$$

$$\dot{\phi} = \Omega_r^x + \tan\theta(\Omega_r^y \sin\phi + \Omega_r^z \cos\phi)$$
$$\dot{\theta} = \Omega_r^y \cos\phi - \Omega_r^z \sin\phi \qquad (58)$$
$$\dot{\psi} = \frac{1}{\cos\theta}(\Omega_r^y \sin\phi + \Omega_r^z \cos\phi)$$

where $\boldsymbol{\Omega} = \begin{bmatrix} \Omega^x & \Omega^y & \Omega^z \end{bmatrix}^T$ is the angular velocity of the spacecraft expressed in the spacecraft's body-fixed frame $S_B$, and $\boldsymbol{\Omega}_r = \begin{bmatrix} \Omega_r^x & \Omega_r^y & \Omega_r^z \end{bmatrix}^T$ is the relative angular velocity of the spacecraft with respect to the orbital frame $S_O$ expressed in the body-fixed frame $S_B$. $\boldsymbol{\Omega}_r$ can be calculated by

$$\boldsymbol{\Omega}_r = \boldsymbol{\Omega} - \boldsymbol{L}_{BO}\boldsymbol{\Omega}_{Orbit} = \boldsymbol{\Omega} - \boldsymbol{L}_{BO}\begin{bmatrix} 0 & -\omega_T & 0 \end{bmatrix}^T \qquad (59)$$

where $\boldsymbol{\Omega}_{Orbit}$ is the angular velocity of the orbital frame $S_O$ expressed in itself.

With the explicit formulations of the GG torque given in APPENDIX, the system of differential equations governing the attitude motion of the spacecraft is autonomous. Numerical simulations can be performed.

The parameters of the asteroid $P$ and its gravity field are assumed to be as follows: $M = 1.4091 \times 10^{12}\,\text{kg}$, $\tau_0 = -7.275 \times 10^4\,\text{m}^2$, $\tau_2 = 1.263 \times 10^4\,\text{m}^2$, and $\omega_T = 1.7453 \times 10^{-4}\,\text{s}^{-1}$. Equation (53) is satisfied by these parameters. The radius of the stationary orbit $R_S$ is equal to 1454.952m by Equation (54).

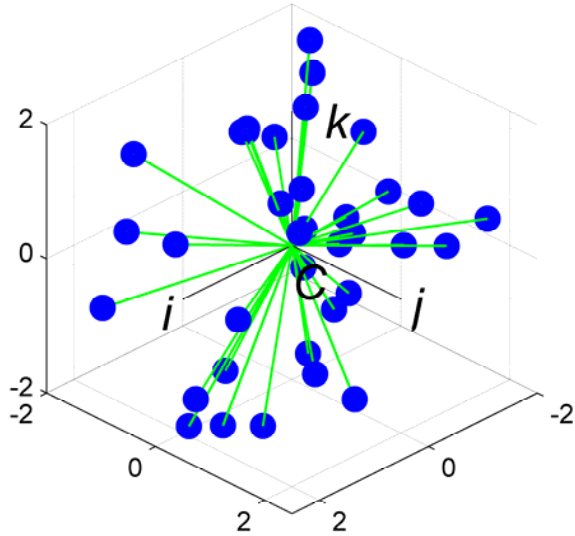

**Figure 4. A Special Spacecraft Consisted of 36 Point Masses Connected by Rigid Massless Rods.**

Here we consider a special spacecraft that is consisted of 36 point masses connected by rigid massless rods, as shown by Figure 4. The mass of each point mass is assumed to be 100kg, and the unit of length in Figure 4 is meter. With the position of every point mass in the body-fixed frame $S_B$ already known, the inertia integrals of the spacecraft can be calculated easily through Equation (36). The gravitational force of each point mass can be calculated by using Equations (15)-(17), and then the exact GG torque of the spacecraft can be obtained by adding the gravitational torque of each point mass with respect to the mass center $C$. Thus we can make comparisons between the motions of previous fourth-order GG torque model, our full fourth-order GG torque model and the exact GG torque. Through these comparisons, different approximate models can be evaluated.

The initial conditions of the numerical simulation are set as that the Euler angles $\psi$, $\theta$ and $\phi$ are all zero, and the spacecraft has the same angular velocity as the orbital reference frame, i.e. $\mathbf{\Omega}_r = \mathbf{0}$. The time histories of the yaw, pitch and roll motions of the spacecraft are given in Figures 5, 6 and 7 respectively. Our full fourth-order GG torque model is denoted by *FourthOrder* in these figures, and previous fourth-order GG torque model is denoted by *PreFourthOrder*. *SecondOrder* is the second-order part of the fourth-order model, i.e. the usual GG torque model in the traditional spacecraft attitude dynamics. And *Precise* is the exact GG torque.

In the case of previous fourth-order GG torque model, the Euler angles $\psi$, $\theta$ and $\phi$ are staying at zero, i.e. the spacecraft is at an equilibrium attitude, just as in the case of the second-order model. That is to say, the previous fourth-order GG torque model has the same equilibria as the second-order model, and the non-central component of the gravity field has no effects on locations of the equilibria in previous fourth-order model.

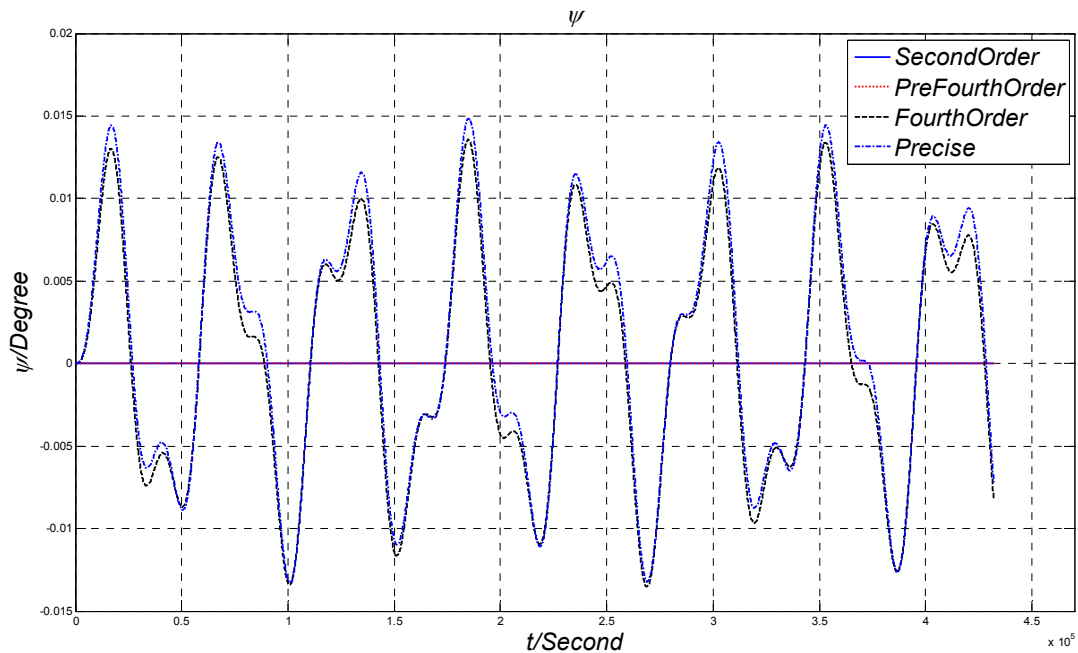

**Figure 5. Yaw Motion of the Spacecraft with Three GG Torque Models and Exact GG Torque.**

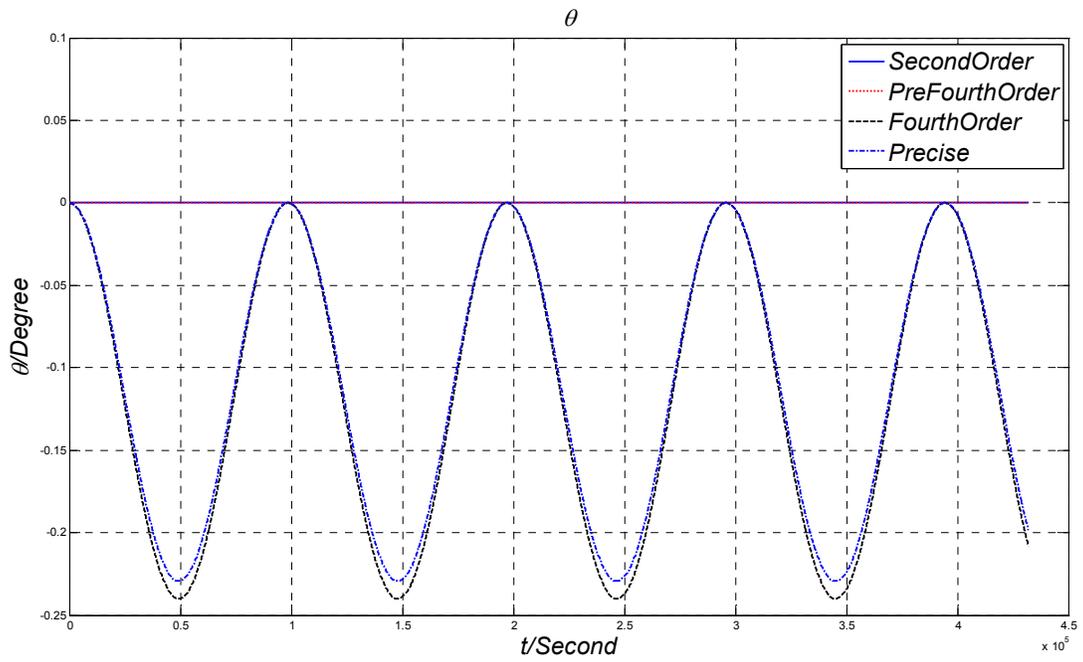

**Figure 6. Pitch Motion of the Spacecraft with Three GG Torque Models and Exact GG Torque.**

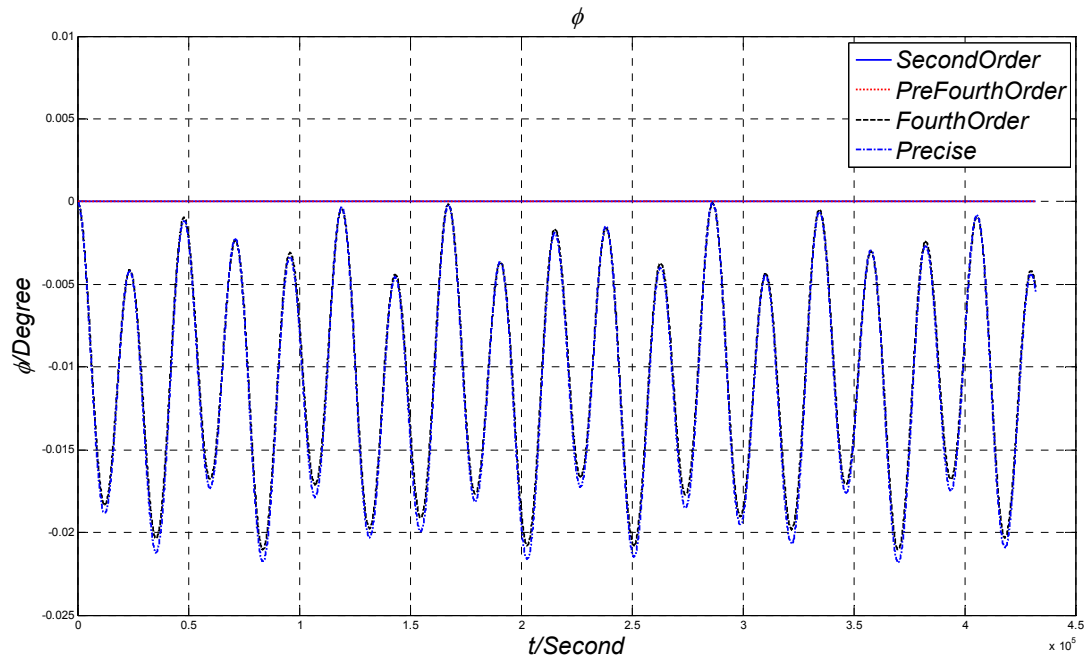

**Figure 7. Roll Motion of the Spacecraft with Three GG Torque Models and Exact GG Torque.**

However, the exact motion of the spacecraft, as shown by *Precise* in Figures 5, 6 and 7, is small amplitude oscillation in all three axes, which is quite different from the previous fourth-order GG torque model. The amplitudes of the yaw and roll motions are the order of $10^{-2}$ degree, while the amplitude of the pitch motion is the order of $10^{-1}$ degree that would be a matter in the high-precise attitude dynamics and control. Therefore, the previous fourth-order GG torque model can not model the attitude motion very well.

From Figures 5, 6 and 7, we can see that our full fourth-order GG torque model fits the exact motion very well with the maximum error order of $10^{-3}$ degree in yaw and roll motions, and $10^{-2}$ degree in the pitch motion. Moreover, our full fourth-order model has the similar equilibria to the exact motion. Therefore, our full fourth-order model is more precise than previous fourth-order model, and is precise enough for high-precise applications in attitude dynamics and control.

**CONCLUSION**

In this paper, a full fourth-order GG torque model of a spacecraft around a non-spherical asteroid is established. In this model, the gravity field of the asteroid is assumed to be 2nd degree and order with harmonic coefficients $C_{20}$ and $C_{22}$. The inertia integrals of the spacecraft up to the fourth-order are considered in our model, which is an improvement with respect to previous fourth-order GG torque model. In previous fourth-order model, inertia integrals of the spacecraft only up to the second-order were considered.

Through Taylor expansion, the mutual gravitational potential up to the fourth-order is derived. Then the explicit formulations of the GG torque of the spacecraft are obtained through the gravitational potential derivatives. We find that the third and fourth-order inertia integrals of the spacecraft appear in the third and fourth-order terms of the GG torque respectively along with the mass of the asteroid. In the previous fourth-order model, the third and fourth-order inertia integrals of the spacecraft were not considered, thus the third-order terms and parts of fourth-order terms of the GG torque were neglected.

A numerical simulation is carried out to verify our full fourth-order GG torque model. In the numerical simulation, a special spacecraft consisted of 36 point masses whose exact motion can be calculated is considered. Simulation results show that the motion of previous fourth-order GG torque model is quite different from the exact motion, while our full fourth-order model fits the exact motion very well. And our full fourth-order model has similar equilibria to the exact motion.

We can conclude that our full fourth-order GG torque model is more sound and precise than previous fourth-order model, and our model is precise enough for high-precise applications in attitude dynamics and control around asteroids.

**ACKNOWLEDGMENTS**

This work is supported by the National Natural Science Foundation of China (ref. 10802002).

**REFERENCES**

[1] K.D. Kumar, "Attitude Dynamics and Control of Satellites Orbiting Rotating Asteroids." *Acta Mechanica*. Vol. 198, 2008, pp. 99–118.

[2] J.-L. Riverin, A.K. Misra, "Attitude Dynamics of Satellites Orbiting Small Bodies." Paper presented at the *AIAA/AAS Astrodynamics Specialist Conference and Exhibit*. Monterey, California, AIAA 2002–4520, 5–8 August 2002.

[3] V.A. Sarychev, "Effects of Earth's Oblateness on Rotational Motion of an Artificial Earth Satellite." *ARS Journal*. 1962, pp. 834–838.

[4] L.B. Schlegel, "Contribution of Earth Oblateness to Gravity Torque on a Satellite." *AIAA Journal*. Vol. 4, 1966, pp. 2075–2077.

[5] P.C. Hughes, *Spacecraft Attitude Dynamics*. New York: John Wiley, 1986, pp. 233–248.

[6] A.J. Maciejewski, "Reduction, Relative Equilibria and Potential in the Two Rigid Bodies Problem." *Celestial Mechanics and Dynamical Astronomy.* Vol. 63, 1995, pp. 1–28.

## APPENDIX: EXPLICIT FORMULATIONS OF $\tilde{T}_B$

The explicit formulations of the full fourth-order GG torque of the spacecraft $\tilde{T}_B$ are given as follows.

$$\begin{aligned}
\tilde{T}_B^x = &\frac{3\mu}{R^3} \bar{R}^y \bar{R}^z (I_{zz} - I_{yy}) + \frac{3\mu}{2R^4} \Big[ \bar{R}^z \left(1 - 5(\bar{R}^y)^2\right) J_{yyy} - 10 \bar{R}^x \bar{R}^y \bar{R}^z J_{xyy} + 10 \bar{R}^x \bar{R}^y \bar{R}^z J_{xzz} \\
&+ \bar{R}^y \left(5(\bar{R}^x)^2 - 1\right) J_{xxz} + \bar{R}^z \left(1 - 5(\bar{R}^z)^2 + 10(\bar{R}^y)^2\right) J_{yzz} + \bar{R}^y \left(5(\bar{R}^z)^2 - 1\right) J_{zzz} \\
&- \bar{R}^y \left(1 - 5(\bar{R}^y)^2 + 10(\bar{R}^z)^2\right) J_{yyz} + 10 \bar{R}^x \left((\bar{R}^y)^2 - (\bar{R}^z)^2\right) J_{xyz} + \bar{R}^z \left(1 - 5(\bar{R}^x)^2\right) J_{xxy} \Big] \\
&+ \frac{5\mu}{2R^5} \Big[ 3 \bar{R}^x \bar{R}^y \left(1 - 7(\bar{R}^z)^2\right) J_{xzzz} + \bar{R}^y \bar{R}^z \left(3 - 7(\bar{R}^z)^2\right) J_{zzzz} + 3 \bar{R}^y \bar{R}^z \left(1 - 7(\bar{R}^x)^2\right) J_{xxzz} \\
&+ 21 \bar{R}^y \bar{R}^z \left((\bar{R}^z)^2 - (\bar{R}^y)^2\right) J_{yyzz} + 3 \bar{R}^y \bar{R}^z \left(7(\bar{R}^x)^2 - 1\right) J_{xxyy} + \bar{R}^x \bar{R}^y \left(3 - 7(\bar{R}^x)^2\right) J_{xxxz} \\
&+ \left(3(\bar{R}^y)^2 - 3(\bar{R}^z)^2 - 7(\bar{R}^y)^4 + 21(\bar{R}^y)^2(\bar{R}^z)^2\right) J_{yyyz} + \bar{R}^y \bar{R}^z \left(7(\bar{R}^y)^2 - 3\right) J_{yyyy} \\
&+ \left(3(\bar{R}^y)^2 - 3(\bar{R}^z)^2 + 7(\bar{R}^z)^4 - 21(\bar{R}^y)^2(\bar{R}^z)^2\right) J_{yzzz} + \bar{R}^x \bar{R}^z \left(7(\bar{R}^x)^2 - 3\right) J_{xxxy} \\
&+ 3 \left((\bar{R}^y)^2 - (\bar{R}^z)^2 + 7(\bar{R}^x)^2(\bar{R}^z)^2 - 7(\bar{R}^x)^2(\bar{R}^y)^2\right) J_{xxyz} + 3 \bar{R}^x \bar{R}^z \left(7(\bar{R}^y)^2 - 1\right) J_{xyyy} \\
&- 3 \bar{R}^x \bar{R}^z \left(1 - 7(\bar{R}^z)^2 + 14(\bar{R}^y)^2\right) J_{xyzz} + 3 \bar{R}^x \bar{R}^y \left(1 - 7(\bar{R}^y)^2 + 14(\bar{R}^z)^2\right) J_{xyyz} \\
&+ \Big( -3\tau_0 \Big( 7 \bar{R}^y \bar{R}^z (\boldsymbol{\gamma} \cdot \boldsymbol{R})^2 - 2 \bar{R}^y \gamma^z (\boldsymbol{\gamma} \cdot \boldsymbol{R}) - 2 \bar{R}^z \gamma^y (\boldsymbol{\gamma} \cdot \boldsymbol{R}) + \frac{2}{5} \gamma^y \gamma^z - \bar{R}^y \bar{R}^z \Big) \\
&- 6\tau_2 \Big( 7 \bar{R}^y \bar{R}^z (\boldsymbol{\alpha} \cdot \boldsymbol{R})^2 - 2 \bar{R}^y \alpha^z (\boldsymbol{\alpha} \cdot \boldsymbol{R}) - 2 \bar{R}^z \alpha^y (\boldsymbol{\alpha} \cdot \boldsymbol{R}) + \frac{2}{5} \alpha^y \alpha^z \Big) \\
&+ 6\tau_2 \Big( 7 \bar{R}^y \bar{R}^z (\boldsymbol{\beta} \cdot \boldsymbol{R})^2 - 2 \bar{R}^y \beta^z (\boldsymbol{\beta} \cdot \boldsymbol{R}) - 2 \bar{R}^z \beta^y (\boldsymbol{\beta} \cdot \boldsymbol{R}) + \frac{2}{5} \beta^y \beta^z \Big) \Big) I_{yy} \\
&+ \Big( 3\tau_0 \Big( 7 \bar{R}^y \bar{R}^z (\boldsymbol{\gamma} \cdot \boldsymbol{R})^2 - 2 \bar{R}^y \gamma^z (\boldsymbol{\gamma} \cdot \boldsymbol{R}) - 2 \bar{R}^z \gamma^y (\boldsymbol{\gamma} \cdot \boldsymbol{R}) + \frac{2}{5} \gamma^y \gamma^z - \bar{R}^y \bar{R}^z \Big) \\
&+ 6\tau_2 \Big( 7 \bar{R}^y \bar{R}^z (\boldsymbol{\alpha} \cdot \boldsymbol{R})^2 - 2 \bar{R}^y \alpha^z (\boldsymbol{\alpha} \cdot \boldsymbol{R}) - 2 \bar{R}^z \alpha^y (\boldsymbol{\alpha} \cdot \boldsymbol{R}) + \frac{2}{5} \alpha^y \alpha^z \Big) \\
&- 6\tau_2 \Big( 7 \bar{R}^y \bar{R}^z (\boldsymbol{\beta} \cdot \boldsymbol{R})^2 - 2 \bar{R}^y \beta^z (\boldsymbol{\beta} \cdot \boldsymbol{R}) - 2 \bar{R}^z \beta^y (\boldsymbol{\beta} \cdot \boldsymbol{R}) + \frac{2}{5} \beta^y \beta^z \Big) \Big) I_{zz} \Big] \qquad (A.1)
\end{aligned}$$

$$\begin{aligned}
\tilde{T}_B^y =& \frac{3\mu}{R^3}\bar{R}^x\bar{R}^z\left(I_{xx}-I_{zz}\right)+\frac{3\mu}{2R^4}\bigg[\bar{R}^z\left(5\left(\bar{R}^x\right)^2-1\right)J_{xxx}+10\bar{R}^x\bar{R}^y\bar{R}^z J_{xxy}-10\bar{R}^x\bar{R}^y\bar{R}^z J_{yzz}\\
&+\bar{R}^z\left(5\left(\bar{R}^y\right)^2-1\right)J_{xyy}+\bar{R}^x\left(1-5\left(\bar{R}^x\right)^2+10\left(\bar{R}^z\right)^2\right)J_{xxz}+\bar{R}^x\left(1-5\left(\bar{R}^z\right)^2\right)J_{zzz}\\
&-\bar{R}^z\left(1-5\left(\bar{R}^z\right)^2+10\left(\bar{R}^x\right)^2\right)J_{xzz}+10\bar{R}^y\left(\left(\bar{R}^z\right)^2-\left(\bar{R}^x\right)^2\right)J_{xyz}+\bar{R}^x\left(1-5\left(\bar{R}^y\right)^2\right)J_{yyz}\bigg]\\
&+\frac{5\mu}{2R^5}\bigg[3\bar{R}^y\bar{R}^z\left(1-7\left(\bar{R}^x\right)^2\right)J_{xxxy}+\bar{R}^x\bar{R}^z\left(3-7\left(\bar{R}^x\right)^2\right)J_{xxxx}+3\bar{R}^x\bar{R}^z\left(1-7\left(\bar{R}^y\right)^2\right)J_{xxyy}\\
&+21\bar{R}^x\bar{R}^z\left(\left(\bar{R}^x\right)^2-\left(\bar{R}^z\right)^2\right)J_{xxzz}+3\bar{R}^x\bar{R}^z\left(7\left(\bar{R}^y\right)^2-1\right)J_{yyzz}+\bar{R}^y\bar{R}^z\left(3-7\left(\bar{R}^y\right)^2\right)J_{xyyy}\\
&+\left(3\left(\bar{R}^z\right)^2-3\left(\bar{R}^x\right)^2-7\left(\bar{R}^z\right)^4+21\left(\bar{R}^x\right)^2\left(\bar{R}^z\right)^2\right)J_{xzzz}+\bar{R}^x\bar{R}^y\left(7\left(\bar{R}^y\right)^2-3\right)J_{yyyz}\\
&+\left(3\left(\bar{R}^z\right)^2-3\left(\bar{R}^x\right)^2+7\left(\bar{R}^x\right)^4-21\left(\bar{R}^x\right)^2\left(\bar{R}^z\right)^2\right)J_{xxxz}+\bar{R}^x\bar{R}^z\left(7\left(\bar{R}^z\right)^2-3\right)J_{zzzz}\\
&+3\left(\left(\bar{R}^z\right)^2-\left(\bar{R}^x\right)^2+7\left(\bar{R}^x\right)^2\left(\bar{R}^y\right)^2-7\left(\bar{R}^y\right)^2\left(\bar{R}^z\right)^2\right)J_{xyyz}+3\bar{R}^x\bar{R}^y\left(7\left(\bar{R}^z\right)^2-1\right)J_{yzzz}\\
&-3\bar{R}^x\bar{R}^y\left(1-7\left(\bar{R}^x\right)^2+14\left(\bar{R}^z\right)^2\right)J_{xxyz}+3\bar{R}^y\bar{R}^z\left(1-7\left(\bar{R}^z\right)^2+14\left(\bar{R}^x\right)^2\right)J_{xyzz}\\
&+\bigg(-3\tau_0\bigg(7\bar{R}^x\bar{R}^z\left(\boldsymbol{\gamma}\cdot\bar{\boldsymbol{R}}\right)^2-2\bar{R}^x\gamma^z\left(\boldsymbol{\gamma}\cdot\bar{\boldsymbol{R}}\right)-2\bar{R}^z\gamma^x\left(\boldsymbol{\gamma}\cdot\bar{\boldsymbol{R}}\right)+\frac{2}{5}\gamma^x\gamma^z-\bar{R}^x\bar{R}^z\bigg)\\
&-6\tau_2\bigg(7\bar{R}^x\bar{R}^z\left(\boldsymbol{\alpha}\cdot\bar{\boldsymbol{R}}\right)^2-2\bar{R}^x\alpha^z\left(\boldsymbol{\alpha}\cdot\bar{\boldsymbol{R}}\right)-2\bar{R}^z\alpha^x\left(\boldsymbol{\alpha}\cdot\bar{\boldsymbol{R}}\right)+\frac{2}{5}\alpha^x\alpha^z\bigg)\\
&+6\tau_2\bigg(7\bar{R}^x\bar{R}^z\left(\boldsymbol{\beta}\cdot\bar{\boldsymbol{R}}\right)^2-2\bar{R}^x\beta^z\left(\boldsymbol{\beta}\cdot\bar{\boldsymbol{R}}\right)-2\bar{R}^z\beta^x\left(\boldsymbol{\beta}\cdot\bar{\boldsymbol{R}}\right)+\frac{2}{5}\beta^x\beta^z\bigg)\bigg)I_{zz}\\
&+\bigg(3\tau_0\bigg(7\bar{R}^x\bar{R}^z\left(\boldsymbol{\gamma}\cdot\bar{\boldsymbol{R}}\right)^2-2\bar{R}^x\gamma^z\left(\boldsymbol{\gamma}\cdot\bar{\boldsymbol{R}}\right)-2\bar{R}^z\gamma^x\left(\boldsymbol{\gamma}\cdot\bar{\boldsymbol{R}}\right)+\frac{2}{5}\gamma^x\gamma^z-\bar{R}^x\bar{R}^z\bigg)\\
&+6\tau_2\bigg(7\bar{R}^x\bar{R}^z\left(\boldsymbol{\alpha}\cdot\bar{\boldsymbol{R}}\right)^2-2\bar{R}^x\alpha^z\left(\boldsymbol{\alpha}\cdot\bar{\boldsymbol{R}}\right)-2\bar{R}^z\alpha^x\left(\boldsymbol{\alpha}\cdot\bar{\boldsymbol{R}}\right)+\frac{2}{5}\alpha^x\alpha^z\bigg)\\
&-6\tau_2\bigg(7\bar{R}^x\bar{R}^z\left(\boldsymbol{\beta}\cdot\bar{\boldsymbol{R}}\right)^2-2\bar{R}^x\beta^z\left(\boldsymbol{\beta}\cdot\bar{\boldsymbol{R}}\right)-2\bar{R}^z\beta^x\left(\boldsymbol{\beta}\cdot\bar{\boldsymbol{R}}\right)+\frac{2}{5}\beta^x\beta^z\bigg)\bigg)I_{xx}\bigg] \quad\text{(A.2)}
\end{aligned}$$

$$\begin{aligned}
\tilde{T}_B^z =& \frac{3\mu}{R^3}\bar{R}^x\bar{R}^y\left(I_{yy}-I_{xx}\right)+\frac{3\mu}{2R^4}\bigg[\bar{R}^y\left(1-5\left(\bar{R}^x\right)^2\right)J_{xxx}+10\bar{R}^x\bar{R}^y\bar{R}^z J_{yyz}-10\bar{R}^x\bar{R}^y\bar{R}^z J_{xxz}\\
&+\bar{R}^x\left(5\left(\bar{R}^z\right)^2-1\right)J_{yzz}+\bar{R}^y\left(1-5\left(\bar{R}^y\right)^2+10\left(\bar{R}^x\right)^2\right)J_{xxy}+\bar{R}^x\left(5\left(\bar{R}^y\right)^2-1\right)J_{yyy}\\
&-\bar{R}^x\left(1-5\left(\bar{R}^x\right)^2+10\left(\bar{R}^y\right)^2\right)J_{xyy}+10\bar{R}^z\left(\left(\bar{R}^x\right)^2-\left(\bar{R}^y\right)^2\right)J_{xyz}+\bar{R}^y\left(1-5\left(\bar{R}^z\right)^2\right)J_{xzz}\bigg]\\
&+\frac{5\mu}{2R^5}\bigg[3\bar{R}^x\bar{R}^z\left(1-7\left(\bar{R}^y\right)^2\right)J_{yyyz}+\bar{R}^x\bar{R}^y\left(3-7\left(\bar{R}^y\right)^2\right)J_{yyyy}+3\bar{R}^x\bar{R}^y\left(1-7\left(\bar{R}^z\right)^2\right)J_{yyzz}\\
&+21\bar{R}^x\bar{R}^y\left(\left(\bar{R}^y\right)^2-\left(\bar{R}^x\right)^2\right)J_{xxyy}+3\bar{R}^x\bar{R}^y\left(7\left(\bar{R}^z\right)^2-1\right)J_{xxzz}+\bar{R}^y\bar{R}^z\left(7\left(\bar{R}^z\right)^2-3\right)J_{xzzz}
\end{aligned}$$

$$+\left(3\left(\bar{R}^x\right)^2-3\left(\bar{R}^y\right)^2-7\left(\bar{R}^x\right)^4+21\left(\bar{R}^x\right)^2\left(\bar{R}^y\right)^2\right)J_{xxxy}+3\bar{R}^y\bar{R}^z\left(7\left(\bar{R}^x\right)^2-1\right)J_{xxxz}$$

$$+\left(3\left(\bar{R}^x\right)^2-3\left(\bar{R}^y\right)^2+7\left(\bar{R}^y\right)^4-21\left(\bar{R}^x\right)^2\left(\bar{R}^y\right)^2\right)J_{xyyy}+\bar{R}^x\bar{R}^y\left(7\left(\bar{R}^x\right)^2-3\right)J_{xxxx}$$

$$+3\left(\left(\bar{R}^x\right)^2-\left(\bar{R}^y\right)^2+7\left(\bar{R}^y\right)^2\left(\bar{R}^z\right)^2-7\left(\bar{R}^x\right)^2\left(\bar{R}^z\right)^2\right)J_{xyzz}+\bar{R}^x\bar{R}^z\left(3-7\left(\bar{R}^z\right)^2\right)J_{yzzz}$$

$$-3\bar{R}^y\bar{R}^z\left(1-7\left(\bar{R}^y\right)^2+14\left(\bar{R}^x\right)^2\right)J_{xyyz}+3\bar{R}^x\bar{R}^z\left(1-7\left(\bar{R}^x\right)^2+14\left(\bar{R}^y\right)^2\right)J_{xxyz}$$

$$+\left(-3\tau_0\left(7\bar{R}^x\bar{R}^y\left(\boldsymbol{\gamma}\cdot\bar{\boldsymbol{R}}\right)^2-2\bar{R}^y\gamma^x\left(\boldsymbol{\gamma}\cdot\bar{\boldsymbol{R}}\right)-2\bar{R}^x\gamma^y\left(\boldsymbol{\gamma}\cdot\bar{\boldsymbol{R}}\right)+\frac{2}{5}\gamma^x\gamma^y-\bar{R}^x\bar{R}^y\right)\right.$$

$$-6\tau_2\left(7\bar{R}^x\bar{R}^y\left(\boldsymbol{\alpha}\cdot\bar{\boldsymbol{R}}\right)^2-2\bar{R}^y\alpha^x\left(\boldsymbol{\alpha}\cdot\bar{\boldsymbol{R}}\right)-2\bar{R}^x\alpha^y\left(\boldsymbol{\alpha}\cdot\bar{\boldsymbol{R}}\right)+\frac{2}{5}\alpha^x\alpha^y\right)$$

$$\left.+6\tau_2\left(7\bar{R}^x\bar{R}^y\left(\boldsymbol{\beta}\cdot\bar{\boldsymbol{R}}\right)^2-2\bar{R}^y\beta^x\left(\boldsymbol{\beta}\cdot\bar{\boldsymbol{R}}\right)-2\bar{R}^x\beta^y\left(\boldsymbol{\beta}\cdot\bar{\boldsymbol{R}}\right)+\frac{2}{5}\beta^x\beta^y\right)\right)I_{xx}$$

$$+\left(3\tau_0\left(7\bar{R}^x\bar{R}^y\left(\boldsymbol{\gamma}\cdot\bar{\boldsymbol{R}}\right)^2-2\bar{R}^y\gamma^x\left(\boldsymbol{\gamma}\cdot\bar{\boldsymbol{R}}\right)-2\bar{R}^x\gamma^y\left(\boldsymbol{\gamma}\cdot\bar{\boldsymbol{R}}\right)+\frac{2}{5}\gamma^x\gamma^y-\bar{R}^x\bar{R}^y\right)\right.$$

$$+6\tau_2\left(7\bar{R}^x\bar{R}^y\left(\boldsymbol{\alpha}\cdot\bar{\boldsymbol{R}}\right)^2-2\bar{R}^y\alpha^x\left(\boldsymbol{\alpha}\cdot\bar{\boldsymbol{R}}\right)-2\bar{R}^x\alpha^y\left(\boldsymbol{\alpha}\cdot\bar{\boldsymbol{R}}\right)+\frac{2}{5}\alpha^x\alpha^y\right)$$

$$\left.\left.-6\tau_2\left(7\bar{R}^x\bar{R}^y\left(\boldsymbol{\beta}\cdot\bar{\boldsymbol{R}}\right)^2-2\bar{R}^y\beta^x\left(\boldsymbol{\beta}\cdot\bar{\boldsymbol{R}}\right)-2\bar{R}^x\beta^y\left(\boldsymbol{\beta}\cdot\bar{\boldsymbol{R}}\right)+\frac{2}{5}\beta^x\beta^y\right)\right)I_{yy}\right] \quad (A.3)$$